\documentclass[aps,prd,floatfix,showpacs,11pt]{revtex4}

\usepackage{amsmath,amsfonts}
\usepackage[dvips]{graphicx}
\usepackage{epsfig}

\newcommand{\ud}{\mathrm{d}}
\newcommand{\ve}{\varepsilon}

\begin{document}

\title{Scalar field cosmology in the energy phase-space -- unified description
of dynamics}

\author{Marek Szyd{\l}owski}
\email{uoszydlo@cyf-kr.edu.pl}
\affiliation{Astronomical Observatory, Jagiellonian University,
Orla 171, 30-244 Krak{\'o}w, Poland}
\affiliation{Mark Kac Complex Systems Research Centre, Jagiellonian University,
Reymonta 4, 30-059 Krak{\'o}w, Poland}
\author{Orest Hrycyna} 
\email{hrycyna@kul.lublin.pl}
\affiliation{Department of Theoretical Physics, Faculty of Philosophy, 
The John Paul II Catholic University of Lublin, Al. Rac{\l}awickie 14, 20-950
Lublin, Poland}

\date{\today}

\begin{abstract}
In this letter we apply dynamical system methods to study all evolutional paths
admissible for all initial conditions of the FRW cosmological model with a
non-minimally coupled to gravity scalar field and a barotropic fluid. We choose
``energy variables'' as phase variables. We reduce dynamics to a 3-dimensional
dynamical system for an arbitrary potential of the scalar field in the phase
space variables
$(\kappa\dot{\phi}/\sqrt{6}H, \kappa\sqrt{V}/\sqrt{3}H,\kappa\phi/\sqrt{6})$.
After postulating the
potential parameter $\Gamma$ as a function of $\lambda$ (defined as $-V'/V$) we
reduce whole dynamics to a 3-dimensional dynamical system and study evolutional
paths leading to current accelerating expansion. If we restrict the form of 
the potential then we will obtain a 2-dimensional dynamical system. We use the 
dynamical system approach to find a new generic quintessence scenario of 
approaching to the de Sitter attractor which appears only for the case of 
non-vanishing coupling constant.
\end{abstract}


\maketitle

\section{Introduction}
\label{sec:1}
At present a scalar field $\phi$ plays a very important role in cosmological
investigations. The discovery of cosmic acceleration \cite{Riess:1998cb,
Perlmutter:1998np} is a motivation to study dynamical models of dark energy
which can be treated as some alternatives to the cosmological constant (see
\cite{Copeland:2006wr} for review). In this context the simplest dynamical
models involving the scalar field $\phi$ (the quintessence idea
\cite{Ratra:1987rm,Wetterich:1987fm}) minimally coupled to gravity with the
potential $V(\phi)$ are considered to model a time dependent equation of the
state parameter $w=p_{\phi}/\rho_{\phi}$, where
$p_{\phi}=\dot{\phi}^{2}/2-V(\phi)$, $\rho_{\phi}=\dot{\phi}^{2}/2+V(\phi)$ and
dot denotes differentiation with respect to the cosmic time $t$.

On the other hand the unknown nature of dark energy expressed in terms of $w$
can be constrained by a variety of astronomical observations. Recently the WMAP
experiment has published its five-year data and polarization power spectra
\cite{Komatsu:2008hk, Dunkley:2008ie} and new supernovae datasets of Union
compilation has also been published \cite{Kowalski:2008ez}. It is interesting
that some inflationary scenarios, i.e. forms of potential function of the scalar
field can be rejected \cite{Komatsu:2008hk, Dunkley:2008ie}. Furthermore, the
single slow-rolling scalar field with potential $V(\phi)\propto\phi^{2}$ is well
within a $2\sigma$ confidence level region, whereas another scalar field potential $V(\phi)\propto\phi^{4}$ has been excluded on more than a $2\sigma$ confidence level \cite{Komatsu:2008hk, Dunkley:2008ie, Kinney:2008wy, Xia:2008ex}. While the simplest candidate for dark energy seems to be a positive cosmological constant, i.e. the LambdaCDM model is favored by a Bayesian model selection method \cite{Kurek:2007tb, Kurek:2007gr, Szydlowski:2006ay}, such an explanation of the accelerated expansion of the universe as associated with the vacuum energy meets the so-called fine tuning problem \cite{Weinberg:1988cp} and it also suffers from the coincidence problem \cite{Zlatev:1998tr}. Therefore various alternative routes have been proposed like phantom \cite{Caldwell:1999ew, Dabrowski:2003jm} or extended quintessence \cite{Faraoni:2006ik, Faraoni:2000gx, Hrycyna:2007mq, Hrycyna:2007gd, Szydlowski:2008zza}. While the minimally coupled scalar field endowed with a quadratic potential function has a strong motivation in inflationary models its generalizations with a simple non-minimal coupling term $\xi R \phi^{2}$ have been studied \cite{Park:2008hz} in the context of the origin of the canonical inflaton field itself. 
 The physical motivation to study of non-minimally coupled scalar field
could be possible application of this models to inflationary cosmology or to the
present dark energy and has a long history (see for example
\cite{Spokoiny:1984bd,
Salopek:1988qh, Fakir:1992cg, Barvinsky:1994hx, Barvinsky:1998rn, Uzan:1999ch,
Chiba:1999wt, Amendola:1999qq, Perrotta:1999am, Holden:1999hm, Bartolo:1999sq,
Boisseau:2000pr, Gannouji:2006jm, Carloni:2007eu, Bezrukov:2007ep,
Barvinsky:2008ia}).

In the present paper, our aim is to perform a complete study of the global
dynamics, attractor properties and stability of trajectories of both canonical
and phantom scalar fields in the framework of dynamical systems methods.
Recently the dynamical system methods was used to find attractor properties of
phantom scalar field \cite{UrenaLopez:2005zd}. Such an analysis was restricted
to the case of the simplest exponential or quadratic potential functions and
minimally coupled scalar field (see also \cite{UrenaLopez:2007vz}). Our study
will include a non-minimal coupling of the scalar field to the gravity in the
form of the term $\xi R \phi^{2}$ and we will not consider any a priori form of
the potential of the scalar field. We take as phase variables the same variables
used in \cite{Copeland:1997et, delaMacorra:1999ff} in the context of minimally
coupled canonical scalar field. The first attempt to study the minimally coupled
scalar field with a quadratic potential cosmology in these variables was made by
Belinsky {\it et al.} \cite{Belinsky:1985zd}. In this parameterization of the
phase space the normalized variables of the scalar field and its first cosmic
time derivative are used. The third new variable is related to the Hubble
function and the resulting system is a $3-$dimensional autonomous one.
Earlier, a minimally coupled massive scalar field in the closed FRW model
was studied by Starobinsky \cite{Starobinsky:1978} where analytical expression
for the matter dominated and quasi-deSitter stages was derived.

In our approach extended to the case of a non-minimally coupled scalar field we
parameterize the phase space using the energy variables determined by the
Friedmann energy equation. The choice of two variables is the same like in the
Belinsky approach, but we complete the set of variables by a normalized variable
related to the so-called roll parameter $\lambda \propto V'/V$, $'\equiv\ud/\ud
\phi$. Therefore our dynamical analysis is the generalization of the studies
\cite{Zhou:2007xp, Chongchitnan:2007eb} to the case of non-vanishing coupling
constant $\xi$.

\section{Dynamics of non-minimally coupled scalar field cosmology in the energy
phase space}
\label{sec:2}

In this section the dynamical system methods are used to study the dynamics of
the scalar field non-minimally coupled to gravity in the Robertson-Walker
geometry. To make the discussion less complex, we ignore any coupling of the
scalar field to matter and assume that the universe is spatially flat. We choose
as phase space variables the so-called ``energy variables'' introduced in
\cite{Copeland:1997et} which in the case of the minimal coupling $(\xi=0)$
assume the form
\begin{equation}
x \equiv \frac{\kappa\dot{\phi}}{\sqrt{6}H} \quad , \qquad y \equiv
\frac{\kappa\sqrt{V(\phi)}}{\sqrt{3}H} ,
\label{eq:1}
\end{equation}
where $V(\phi)$ is a potential function of the scalar field $\phi$,
$H=\dot{a}/{a}$ is the Hubble parameter and $a$ is the scale factor. These
variables determine the ratio of kinetic and potential energy of the scalar
field to the total energy
\begin{equation}
\rho_{\phi}=\frac{1}{2}\dot{\phi}^{2} + V(\phi) \quad \Rightarrow \quad
\frac{\frac{1}{2}\dot{\phi}^{2}}{\rho_{\phi}} + \frac{V(\phi)}{\rho_{\phi}}=1 
\end{equation}
where $\rho_{\phi}=\frac{3}{\kappa^{2}} H^{2}$ for the flat model filled with the scalar field only. 

If we additionally postulate the presence of matter which satisfies the
barotropic equation of state, then $\rho_{\phi}+\rho_{m}=\frac{3}{\kappa^{2}}
H^{2}$, where $p_{m}=w_{m}\rho_{m}$ and $w_{m}$ is constant and energy
conservation requires 
\begin{equation}
\frac{\kappa^{2}\rho_{\phi}}{3H^{2}}+\frac{\kappa^{2}\rho_{m}}{3H^{2}}= x^{2} + y^{2} + \Omega_{m}
=1 ,
\label{eq:3}
\end{equation}
where $\Omega_{m}$ is the matter density parameter. Because of (\ref{eq:3}) the
trajectories in the phase space will always be located within the unit circle in
the phase space. If we restrict ourselves to positive and monotonically
decreasing potential functions $V(\phi)$ we can consider only the trajectories
located in one quadrant of the energy phase space.

Above considerations can be simply generalized to the case of the non-minimal
coupling of the scalar field (both canonical and phantom) to the curvature. In
this case the action for the scalar field and gravity assumes the following
form 
\begin{equation}
S=\frac{1}{2}\int\ud^{4}x\sqrt{-g}\Bigg(\frac{1}{\kappa^{2}}R -
\varepsilon\Big(g^{\mu\nu}\partial_{\mu}\phi\partial_{\nu}\phi+\xi
R\phi^{2}\Big) - 2U(\phi)\Bigg)
\label{eq:4}
\end{equation}
where $\kappa^{2}=8\pi G$, $\ve=+1,-1$ corresponds to canonical and phantom
scalar fields respectively, the metric signature is $(-,+,+,+)$ and
$R=6(\frac{\ddot{a}}{a}+\frac{\dot{a}^{2}}{a^{2}})$ is the Ricci scalar, $a$ is
the scale factor and a dot denotes differentiation with respect to the
cosmological time.

After dropping the full derivatives with respect to time we obtain the
dynamical equation for scalar field from variation
$\delta S/\delta \phi=0$
\begin{equation}
\ddot{\phi}+3H\dot{\phi} +\xi R \phi + \ve U'(\phi)= 0.
\end{equation}
as well as the energy conservation condition from variation
$\delta S/\delta g=0$
\begin{equation}
\mathcal{E} = \ve\frac{1}{2}\dot{\phi}^{2} + \ve 3\xi H^{2}\phi^{2} + 
\ve 3\xi H (\phi^{2})\dot{} + U(\phi) - \frac{3}{\kappa^{2}}H^{2}
\end{equation}

If we postulate the existence of a barotropic fluid in the universe the
conservation condition reads
\begin{equation}
\frac{3}{\kappa^{2}}H^{2}=\rho_{\phi}+\rho_{m},
\end{equation}
\begin{equation}
\dot{H}=-\frac{\kappa^{2}}{2}\Big((\rho_{\phi}+p_{\phi})+\rho_{m}(1+w_{m})\Big)
\end{equation}
from which we can simply receive the energy density and the pressure of the
scalar field, namely
\begin{subequations}
\begin{eqnarray}
\label{eq:6a}
\rho_{\phi}&=& \varepsilon\frac{1}{2}\dot{\phi}^{2} + U(\phi) +
\varepsilon 6 \xi H\phi\dot{\phi} + \ve3\xi H^{2}\phi^{2},\\
\label{eq:6b}
p_{\phi}&=& \varepsilon\frac{1}{2}(1-4\xi)\dot{\phi}^{2} - U(\phi) + 
\varepsilon 2 \xi
H \phi\dot{\phi} - \varepsilon 2 \xi (1-6\xi)\dot{H}\phi^{2} -
\varepsilon3\xi(1-8\xi)H^{2}\phi^{2} + 2\xi\phi U'(\phi).
\end{eqnarray}
\end{subequations}
and in the flat FRW model the equation of state (EoS) for the scalar field is
given by
\begin{equation}
w_{\phi}=\frac{p_{\phi}}{\rho_{\phi}}.
\end{equation}

In the case of non-minimal coupling $\xi\ne0$ the additional variable $z$ should
be chosen in the parameterization of dynamics
\begin{equation}
z\equiv \frac{\kappa}{\sqrt{6}}\phi.
\end{equation}
The choice of phase variables is suggested as in the case of minimal coupling by
the energy conservation condition written as
\begin{equation}
\frac{\kappa^{2}\rho_{\phi}}{3H^{2}}+\frac{\kappa^{2}\rho_{m}}{3H^{2}}=1
\end{equation}
or in terms of dynamical variables $x$, $y$, $z$
\begin{equation}
\label{eq:11}
\Omega_{\phi}=y^{2}+\ve\big((1-6\xi)x^{2}+6\xi(x+z)^{2}\big)=1-\Omega_{m}
\end{equation}
where the sense of variables $x$ and $y$ is preserved like in $\xi=0$ case. This
condition defines in the phase space a domain $\Omega_{\phi}\ge0$ admissible for
motion. If $\Omega_{m}=0$ then condition (\ref{eq:11}) define a $2$-dimensional 
surface in the space $(x,y,z)$.

Similarly the acceleration equation can be rewritten to the form
\begin{equation}
\dot{H}=-\frac{\kappa^2}{2}(\rho_{\text{eff}}+p_{\text{eff}}) = -\frac{3}{2}H^{2}(1+w_{\text{eff}}),
\label{eq:12}
\end{equation}
where 
\begin{equation}
\label{eq:13}
w_{\text{eff}}=\frac{p_{\text{eff}}}{\rho_{\text{eff}}} = \frac{p_{\phi}+w_{m}\rho_{m}}{\rho_{\text{eff}}} =
\frac{\kappa^{2}p_{\phi}}{3H^{2}}+w_{m}\Omega_{m}
\end{equation}
and after the substitution of (\ref{eq:12}) to (\ref{eq:6b}) expressed in terms
of variables $(x,y,z)$ we obtain
\begin{equation}
\frac{\kappa^{2}p_{\phi}}{3H^{2}}= \ve(1-4\xi)x^{2}-y^{2}(1+2\xi\lambda z) +
\ve4\xi x z + \ve 12 \xi^{2}z^{2} + \ve6\xi(1-6\xi)z^{2}w_{eff}
\end{equation}
where $\lambda\equiv-\frac{\sqrt{6}}{\kappa}\frac{U'(\phi)}{U(\phi)}$. Finally
after substitution of above formula into (\ref{eq:13}) we can explicitly
calculate $w_{\text{eff}}$, namely
\begin{equation}
\label{weff}
w_{\text{eff}}=\frac{\ve(1-4\xi)x^{2}-y^{2}(1+2\xi\lambda z) + \ve4\xi x z + \ve 12
\xi^{2}z^{2} + w_{m}\Omega_{m}}{1-\ve6\xi(1-6\xi)z^{2}},
\end{equation}
where $\Omega_{m}$ is given by equation (\ref{eq:11}).

Let us start to find the dynamical system describing the evolution of our
model---the non-minimally coupled scalar field in the Robertson-Walker background. 
For derivation of basic equations we take $\log{x}$ (or $\log{y}$) and then
differentiate with respect to the cosmic time variable both sides of
the expression. Then we obtain
\begin{equation}
\frac{\dot{x}}{x} = \frac{\ddot{\phi}}{\dot{\phi}} - \frac{\dot{H}}{H} = -3H
-\xi R \frac{\phi}{\dot{\phi}} - \ve \frac{U'(\phi)}{\dot{\phi}} -
\frac{\dot{H}}{H}
\end{equation}
where we have used the dynamical equation for the motion of the scalar field
$\ddot{\phi} + 3H\dot{\phi} + \xi R\phi + \ve U'(\phi)=0$ and
$R=6(\frac{\ddot{a}}{a}+\frac{\dot{a}^{2}}{a^{2}})= 6(\dot{H}+2H^{2})$ is the
Ricci scalar. The above equation reduces to
\begin{equation}
\frac{1}{H}\frac{\dot{x}}{x} = -3 -
6\xi\frac{\kappa}{\sqrt{6}}\frac{\dot{H}+2H^{2}}{H^{2}}\frac{\phi}{x} - \ve
\frac{\kappa}{\sqrt{6}}\frac{U'(\phi)}{H^{2}x} - \frac{\dot{H}}{H^{2}}
\end{equation}
after the reparameterization $\dot{x}=H\frac{\ud x}{\ud \ln{a}}$ we obtain
\begin{equation}
\frac{\ud x}{\ud \ln{a}} = -3x - 12\xi z + \ve \frac{1}{2} \lambda y^{2} -
\frac{\dot{H}}{H^{2}} (x+6\xi z)
\end{equation}
Finally, we obtain
\begin{eqnarray}
\label{sys1}
\frac{\ud x}{\ud \ln{a}} &=& -3x -12 \xi z + \ve \frac{1}{2}\lambda y^{2}+
\nonumber \\ &+&
\frac{3}{2} \frac{x+6\xi z}{1-\ve6\xi(1-6\xi)z^{2}}
\Big\{1-\ve6\xi(1-8\xi)z^{2}+\ve(1-4\xi)x^{2}-y^{2}(1+2\xi\lambda z) + 
\ve4\xi x z + w_{m}\Omega_{m}\Big\}
\end{eqnarray}
where $\Omega_{m}$ is determined from the constraint condition (\ref{eq:11}).

The same method can be adopted to the variable
$y=\frac{\kappa\sqrt{V}}{\sqrt{3}H}$, namely
\begin{equation}
\frac{\dot{y}}{y} = \frac{1}{2}\frac{U'(\phi)}{U(\phi)}\dot{\phi} -
\frac{\dot{H}}{H},
\end{equation}
and
\begin{eqnarray}
\label{sys2}
\frac{\ud y}{\ud \ln{a}} &=& - \frac{1}{2}\lambda x y \nonumber \\ &+& 
\frac{3}{2}
\frac{y}{1-\ve6\xi(1-6\xi)z^{2}}\Big\{1-\ve6\xi(1-8\xi)z^{2}+\ve(1-4\xi)x^{2}-
y^{2}(1+2\xi\lambda z)+\ve4\xi x z + w_{m}\Omega_{m}\Big\}
\end{eqnarray}

And after the elimination of $\Omega_{m}$ equations (\ref{sys1}) and
(\ref{sys2}) can be presented in the form 
\begin{eqnarray}
\label{sys1a}
\frac{\ud x}{\ud \ln{a}} \big(1-\ve6\xi(1-6\xi)z^{2}\big) &=& -3x -12\xi z
+\ve\frac{1}{2}\lambda y^{2}\big(1-\ve6\xi(1-6\xi)\big) +\ve6\xi(1-6\xi)x z^{2}
+ \nonumber \\
& & + \frac{3}{2}(x+6\xi z) \Big(\ve(1-6\xi)(1-w_{m})x^{2} +
\ve2\xi(1-3w_{m})(x+z)^{2} + \nonumber \\ & & \quad \qquad \qquad \qquad +
(1+w_{m})(1-y^{2})\Big)
\end{eqnarray}

\begin{eqnarray}
\label{sys2a}
\frac{\ud y}{\ud \ln{a}} \big(1-\ve6\xi(1-6\xi)z^{2}\big) &=&
-\frac{1}{2}\lambda y \Big( x\big(1-\ve6\xi(1-6\xi)z^{2}\big)+6\xi y^{2}z\Big)
-\ve12\xi(1-6\xi)yz^{2} + \nonumber \\ & & + \frac{3}{2}y
\Big(\ve(1-6\xi)(1-w_{m})x^{2} +
\ve2\xi(1-3w_{m})(x+z)^{2} + \nonumber \\ & & \quad \qquad +
(1+w_{m})(1-y^{2})\Big)
\end{eqnarray}

where $w_{m}=0$, $1/3$ for dust matter and radiation respectively.

The dynamical equations (\ref{sys1a}) and (\ref{sys2a}) should be completed by
two additional equations to make the dynamical system closed, namely
\begin{equation}
\label{sys3a}
\frac{\ud z}{\ud \ln{a}} = x,
\end{equation}
and the last equation can be established from the definition of $\lambda$
variable
\begin{equation}
\label{sys4a}
\frac{\ud \lambda}{\ud \ln{a}} = -\lambda^{2}\big(\Gamma-1\big)x
\end{equation}
where $\Gamma=\frac{U'' U}{U'^{2}}$.

Making following time reparameterization 
\begin{equation}
\big(1-\ve6\xi(1-6\xi)z^{2}\big)\frac{\ud}{\ud \ln{a}} = \frac{\ud}{\ud \tau}
\end{equation}
we can write complete dynamical system in the form
\begin{subequations}
\begin{eqnarray}
x' &=& -3x -12\xi z
+\ve\frac{1}{2}\lambda y^{2}\big(1-\ve6\xi(1-6\xi)\big) +\ve6\xi(1-6\xi)x z^{2}
+ \nonumber \\
& & + \frac{3}{2}(x+6\xi z) \Big(\ve(1-6\xi)(1-w_{m})x^{2} +
\ve2\xi(1-3w_{m})(x+z)^{2} +
(1+w_{m})(1-y^{2})\Big), \\
y' &=& -\frac{1}{2}\lambda y \Big( x\big(1-\ve6\xi(1-6\xi)z^{2}\big)+6\xi
y^{2}z\Big)
-\ve12\xi(1-6\xi)yz^{2} + \nonumber \\ & & + \frac{3}{2}y
\Big(\ve(1-6\xi)(1-w_{m})x^{2} +
\ve2\xi(1-3w_{m})(x+z)^{2} +
(1+w_{m})(1-y^{2})\Big), \\
z' &=& x\big(1-\ve6\xi(1-6\xi)z^{2}\big), \\
\lambda' & = & -\lambda^{2}(\Gamma-1)x\big(1-\ve6\xi(1-6\xi)z^{2}\big)
\end{eqnarray}
\end{subequations}

It is also another way to eliminate one of the variables, namely $z$. If we
assume that $\Gamma=\Gamma(\lambda)$ then the variable $z$ can be expressed by
$\lambda$ according to the formula
\begin{equation}
z=-\int^{\lambda}\frac{\ud \lambda}{\lambda^{2}\big(\Gamma(\lambda)-1\big)},
\end{equation}
If we assume that $\Gamma(\lambda)=1-\alpha/\lambda^{2}$ for an arbitrary
constant $\alpha$, then $z(\lambda)$ can be integrated in the exact form 
$$
z(\lambda)=\frac{\lambda}{\alpha}.
$$

Let us now make some important remarks about the general properties of the
system describing the evolution of the non-minimally coupled scalar field on the
background of the Robertson-Walker symmetry:
\begin{itemize}
\item[1.]{In the general case of non-vanishing barotropic matter the system is
reduced to the form of a $3$-dimensional autonomous dynamical system which can
be studied by the dynamical systems methods. The motion of the system is
restricted to the domain $\Omega_{\phi}(x,y,z)\ge0$}
\item[2.]{If effects of the barotropic matter are not considered then motion of
the system in the phase space is restricted to the $2$-dimensional surface
$\Omega_{\phi}=1$}
\item[3.]{If the form of the potential function is assumed at the very
beginning then the scalar field cosmological model is represented by a
$2$-dimensional dynamical system.}
\end{itemize}

\section{Dynamical system without the matter $\Omega_{m}=0$}

Let us consider the model without matter. The motion of the system takes place 
on the surface determined by equation (11). Putting $\Omega_{m}=0$ this surface
is given by
\begin{equation}
\Omega_{\phi}=1 \quad \Longrightarrow \quad y^{2}=1-\ve\big((1-6\xi)x^{2}+6\xi(x+z)^{2}\big).
\end{equation}
By using this equation we can eliminate the variable $y$ from equation (\ref{weff}) and
now the effective equation of state coefficient assumes the following form 
\begin{equation}
w_{\text{eff}}=\frac{1}{1-\ve6\xi(1-6\xi)z^{2}}\Big\{ \ve(1-4\xi)x^{2}+\ve4\xi x
z + \ve12\xi^{2}z^{2}-(1+2\xi\lambda z)\Big(1-\ve\big((1-6\xi)x^{2}+6\xi(x+z)^{2}\big)\Big)\Big\}
\end{equation}
If we assume that $\Gamma=\Gamma(\lambda)$ then from equation (27) we can
eliminate the variable $\lambda$ and reduce the dynamical system to the two-dimensional
one
\begin{subequations}
\begin{eqnarray}
\frac{\ud x}{\ud \tau} &=& -x\big(1-\ve6\xi(1-6\xi)z^{2}\big) +
\ve(1-6\xi)(x+6\xi z)x^{2} + \nonumber \\ & &+
\Big(1-\ve\big((1-6\xi)x^{2}+6\xi(x+z)^{2}\big)\Big)
\Big(\ve\frac{1}{2}\lambda(z)(1-\ve6\xi z(x+z))-2(x+6\xi z)\Big), \\
\frac{\ud z}{\ud \tau} &=& x\big(1-\ve6\xi(1-6\xi)z^{2}\big).
\end{eqnarray}
\end{subequations}
Dynamics of this system for a quadratic potential function
$\lambda(z)=-2\frac{1}{z}$ and the phantom scalar field $\ve=-1$ have been
studied in our previous work \cite{Hrycyna:2007gd}. 

Another possibility is the elimination of the variable $z$ and then the dynamical
system is in the form
\begin{subequations}
\begin{eqnarray}
\frac{\ud x}{\ud \tau} &=& -x\big(1-\ve6\xi(1-6\xi)z^{2}(\lambda)\big) +
\ve(1-6\xi)(x+6\xi z(\lambda))x^{2} + \nonumber \\ & &+
\Big(1-\ve\big((1-6\xi)x^{2}+6\xi(x+z(\lambda))^{2}\big)\Big)
\Big(\ve\frac{1}{2}\lambda(1-\ve6\xi z(\lambda)(x+z(\lambda)))-2(x+6\xi
z(\lambda))\Big), \\
\frac{\ud \lambda}{\ud \tau} &=& -\lambda^{2}(\Gamma(\lambda)-1)x\big(1-\ve6\xi(1-6\xi)z^{2}\big).
\end{eqnarray}
\end{subequations}

\begin{figure}
\includegraphics[scale=1]{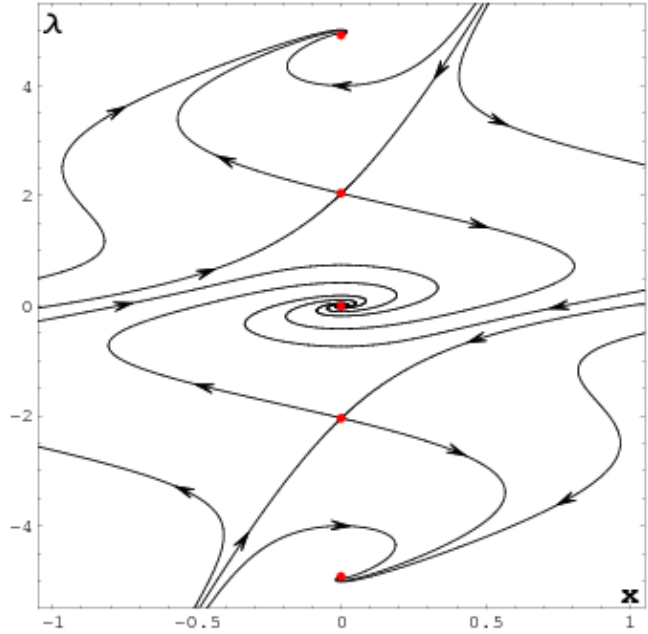}
\caption{Phase portrait on the plane $(x,\lambda)$ in the finite domain for
model without the matter
$\Omega_{m}=0$ and $\Gamma(\lambda)=1-\frac{\alpha}{\lambda^{2}}$ as an
example. 
The critical point in the origin $(0,0)$ is of a stable focus type (see
Table \ref{tab:1}).}
\label{fig:1}
\end{figure}

For the model with $\Omega_{m}=0$ and arbitrary coupling constant we have following
critical points at finite domain of the phase space
\begin{itemize}
\item{($x_{0}=0$, $\lambda_{0}=0:z(\lambda_{0})=0$) \\ 
the effective equation of state parameter calculated at this point is
\begin{equation}
w_{\text{eff}}=-1
\end{equation}
\\
the eigenvalues of the linearization matrix are
\begin{equation}
l_{1}=-\frac{1}{2}(3+\sqrt{9+\ve2\alpha-48\xi}), \qquad
l_{2}=-\frac{1}{2}(3-\sqrt{9+\ve2\alpha-48\xi})
\end{equation}
where $\alpha=f(\lambda_{0})=-\lambda_{0}^{2}(\Gamma(\lambda_{0})-1)$
}
\item{($x_{0}=0$,
$\lambda_{0}=\text{const.}:z^{2}(\lambda_{0})=\frac{1}{\ve6\xi}$) \\
the effective equation of state parameter is
\begin{equation}
w_{\text{eff}}=\frac{1}{3}
\end{equation}
\\
the eigenvalues of the linearization matrix are
\begin{equation}
l_{1}=-6\xi, \qquad l_{2}=24\xi
\end{equation}
}
\end{itemize}

The stability of the critical points depends on the sign of the real part 
of the eigenvalues. The first critical point will be always stable if only 
$9 + 2\alpha \epsilon - 48 \xi < 0$. For the second critical point which 
represents the radiation dominated universe we obtain a saddle critical point. 

\begin{table}
\caption{The simplest finite critical points for model with no matter content
$\Omega_{m}=0$ and arbitrary coupling constant $\xi$.}
\label{tab:1}
\begin{tabular}{|c|c|c|c|c|}
\hline
& Critical point & $w_{\text{eff}}$ & eigenvalues & existence \\
\hline
1) & $x_{0}=0$, $\lambda_{0}=0$ & $-1$ &
$l_{1}=-\frac{1}{2}(3+\sqrt{9+\ve2\alpha-48\xi})$ & $z(\lambda_{0})=0$\\
& & & $l_{2}=-\frac{1}{2}(3-\sqrt{9+\ve2\alpha-48\xi})$ & \\
\hline
2) & $x_{0}=0$, $\lambda_{0}=\text{const}$ &
$\frac{1}{3}$ & $l_{1}=-6\xi$ & $\ve\xi>0$ \\
& & & $l_{2}=24\xi$ & $z^{2}(\lambda_{0})=\frac{1}{\ve6\xi}$\\
\hline
\end{tabular}
\end{table}

\section{Model with the matter $\Omega_{m}\ne0$}

Let us consider the general case of the model with barotropic matter and 
the scalar field with non-minimal coupling. The evolution of the model 
is represented by a 4-dimensional dynamical system with polynomial right-hand 
sides
\begin{equation}
\Omega_{m} =  1-y^{2}-\ve\big((1-6\xi)x^{2}+6\xi(x+z)^{2}\big)
\end{equation}
and effective equation of state parameter in terms of dynamical variables reads
\begin{align}
w_{\text{eff}} = \frac{1}{1-\ve6\xi(1-6\xi)z^{2}}\Big\{& -1 +
\ve(1-6\xi)(1-w_{m})x^{2} + \ve2\xi(1-3w_{m})(x+z)^{2} + \nonumber \\ 
& + (1+w_{m})(1-y^{2}) - \ve2\xi(1-6\xi)z^{2}-2\xi\lambda y^{2}z\Big\},
\label{weff_nm}
\end{align}

Dynamical system is in the form
\begin{subequations}
\begin{eqnarray}
\frac{\ud x}{\ud \tau} &=& -3x -12\xi z
+\ve\frac{1}{2}\lambda y^{2}\big(1-\ve6\xi(1-6\xi)\big) +\ve6\xi(1-6\xi)x z^{2}
+ \nonumber \\
& & + \frac{3}{2}(x+6\xi z) \Big(\ve(1-6\xi)(1-w_{m})x^{2} +
\ve2\xi(1-3w_{m})(x+z)^{2} +
(1+w_{m})(1-y^{2})\Big), \\
\frac{\ud y}{\ud \tau} &=& -\frac{1}{2}\lambda y \Big( x\big(1-\ve6\xi(1-6\xi)z^{2}\big)+6\xi
y^{2}z\Big)
-\ve12\xi(1-6\xi)yz^{2} + \nonumber \\ & & + \frac{3}{2}y
\Big(\ve(1-6\xi)(1-w_{m})x^{2} +
\ve2\xi(1-3w_{m})(x+z)^{2} +
(1+w_{m})(1-y^{2})\Big), \\
\frac{\ud z}{\ud \tau} &=& x\big(1-\ve6\xi(1-6\xi)z^{2}\big), \\
\frac{\ud \lambda}{\ud \tau} & = & -\lambda^{2}(\Gamma-1)x\big(1-\ve6\xi(1-6\xi)z^{2}\big)
\end{eqnarray}
\label{system1}
\end{subequations}

In the special case of a minimally coupled scalar field the dynamical system is reduced to
the following
\begin{subequations}
\begin{eqnarray}
\frac{\ud x}{\ud \ln{a}} &=& -3x +\ve\frac{1}{2}\lambda y^{2} +
\frac{3}{2}x\Big(\ve(1-w_{m})x^{2}+(1+w_{m})(1-y^{2})\Big), \\
\frac{\ud y}{\ud \ln{a}} &=& -\frac{1}{2}\lambda x y
+\frac{3}{2}y\Big(\ve(1-w_{m})x^{2}+(1+w_{m})(1-y^{2})\Big), \\
\frac{\ud z}{\ud \ln{a}} &=& x,\\
\frac{\ud \lambda}{\ud \ln{a}} &=& -\lambda^{2}\big(\Gamma-1\big)x.
\end{eqnarray}
\end{subequations}

In what follows we will assume that $\Gamma=\Gamma(\lambda)$ and that for any
critical point exists $\lambda_{0}$ such that
\begin{equation}
\label{eq:c1}
f(\lambda_{0})=\lambda_{0}^{2}(\Gamma(\lambda_{0})-1)=-\alpha
\end{equation}
and
\begin{equation}
\label{eq:c2}
\frac{\ud f(\lambda)}{\ud \lambda}|_{\lambda_{0}}=f'(\lambda_{0})=\text{const}
\end{equation}
are finite.

For non-minimally coupled scalar field in the energy phase space we have
following critical points:
\begin{itemize}
\item{$x_{0}=0$, $y_{0}=0$, and $\lambda_{0}=\text{const.}$ where $\lambda_{0}$
is an arbitrary constant such that $z(\lambda_{0})=0$:

the effective equation of state parameter calculated at this point is
\begin{equation}
w_{\text{eff}}=w_{m}
\end{equation}
the eigenvalues of the linearization matrix are
\begin{subequations}
\begin{eqnarray}
l_{1}&=&-\frac{3}{4}(1-w_{m})\Big(1+\sqrt{1+\frac{16}{3}\xi\frac{1-3w_{m}}{(1-w_{m})^{2}}}\Big),
\nonumber \\
l_{2}&=&\frac{3}{2}(1+w_{m}), \\
l_{3}&=&-\frac{3}{4}(1-w_{m})
\Big(1-\sqrt{1+\frac{16}{3}\xi\frac{1-3w_{m}}{(1-w_{m})^{2}}}\Big)
\nonumber
\end{eqnarray}
\end{subequations}

For barotropic fluid equation of state parameter $w_{m}>-1$ the eigenvalue $l_{2}$ is always positive and this critical
point corresponds to an unstable focus when the eigenvalues $l_{1}$ and
$l_{3}$ are
complex numbers or to a saddle when $l_{1}$ and $l_{3}$ are purely real and
negative. In
opposite case, when $w_{m}<-1$ this point correspond to a sink (i.e. a focus or
a stable node, depending on the value of square root in $l_{1}$ and $l_{3}$)
because the real parts of the eigenvalues are always negative.
}

\item{$(x_{0}=0$, $y_{0}^{2}=1$, $\lambda_{0}=0:z(\lambda_{0})=0)$ 

the effective equation of state parameter at this point is
\begin{equation}
w_{\text{eff}}=-1
\end{equation}
the simplest form of the
function $\Gamma(\lambda)$ which fulfills both conditions (\ref{eq:c1}) and
(\ref{eq:c2}) is 
\begin{equation}
\Gamma(\lambda)=1-\frac{\alpha}{\lambda^{2}}
\end{equation}
the eigenvalues of the linearization matrix are
\begin{equation}
l_{1}=-\frac{1}{2}(3+\sqrt{9+\ve2\alpha-48\xi}), \quad l_{2}=-3(1+w_{m}), \quad
l_{3}=-\frac{1}{2}(3-\sqrt{9+\ve2\alpha-48\xi})
\end{equation}
the character of this critical point depends on the value of $\alpha$ and $\xi$.
}
\item{($x_{0}=0$, $y_{0}=0$, $\lambda_{0}=\text{const} :
z^{2}(\lambda_{0})=\frac{1}{\ve6\xi}$)
the effective equation of state parameter is
\begin{equation}
w_{\text{eff}}=\frac{1}{3}
\end{equation}
and the eigenvalues of linearization matrix are
\begin{equation}
l_{1}=-6\xi, \qquad l_{2}=12\xi, \qquad l_{3}=6\xi(1-3w_{m})
\end{equation}

At this critical point the effective gravitational constant changes sign an
any FRW model becomes unstable with respect to arbitrary small anisotropic or
inhomogeneous perturbations and a curvature singularity forms in this point as
it was shown first by Starobinsky \cite{Starobinsky:1981}.
}

\end{itemize}

\begin{figure}
\includegraphics[scale=1]{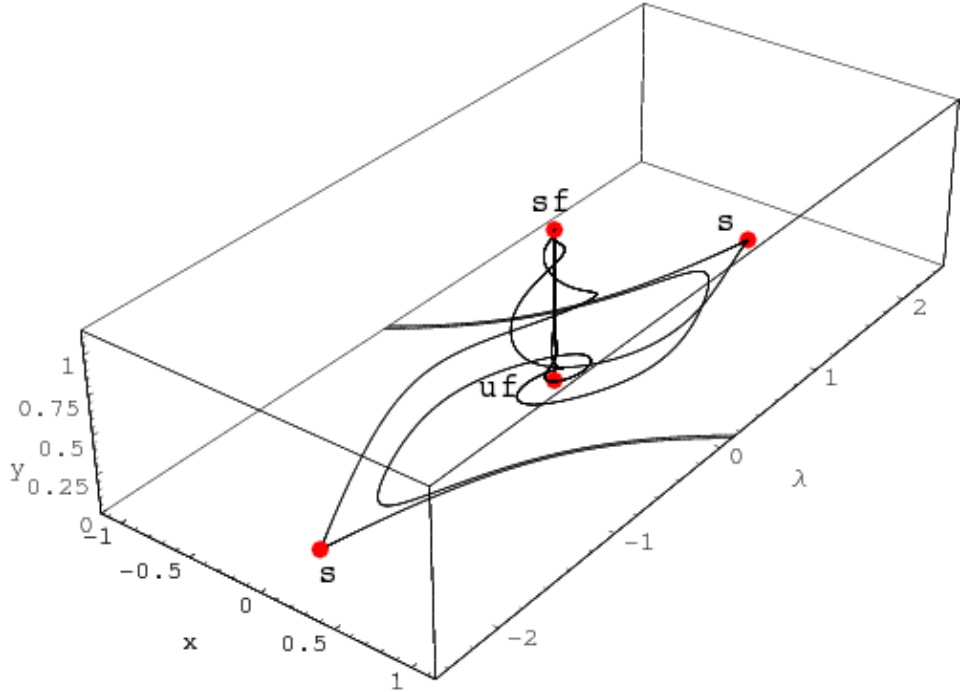}
\caption{Three dimensional phase portrait of the system (\ref{system1}). For
illustration we put $\Gamma(\lambda)=1-\frac{\alpha}{\lambda^{2}}$ and the same
values of $\xi$ and $\alpha$ as in Fig.~\ref{fig:1}. The figure illustrates a
quintessence multiple scenario with a stable focus type critical point $sf$ as a
final state which is the deSitter attractor ($w_{\text{eff}}=-1$) and
a saddle type critical point $s$ ($w_{\text{eff}}=1/3$) and an unstable focus
$uf$ ($w_{\text{eff}}=w_{m}$) as intermediate states.}
\label{fig:2}
\end{figure}

\begin{figure}
\includegraphics[scale=1]{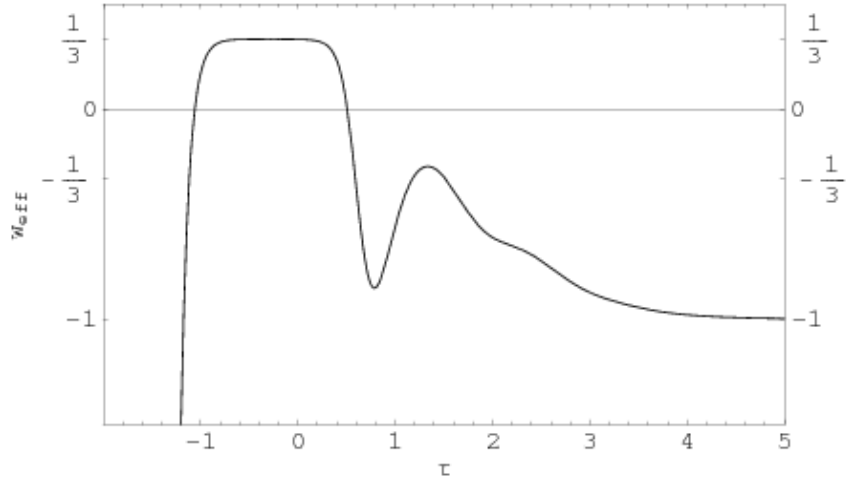}
\caption{Plot of the evolution of $w_{\text{eff}}$ (the relation
(\ref{weff_nm})) for the non-minimally
coupled canonical scalar field and positive coupling constant $\xi$. The probing
trajectory used to plot this relation starts its evolution at $\tau_{0}=0$ near
a saddle type critical point ($w_{\text{eff}}=1/3$) and then approaches an
unstable focus 
critical point $w_{\text{eff}}=w_{m}=-1/3$ and next escapes to the stable
de Sitter critical point with $w_{\text{eff}}=-1$. Note the existence of a short
time interval during which $w_{\text{eff}}\simeq\frac{1}{3}$.}
\label{fig:3}
\end{figure}

\begin{table}
\caption{The simplest finite critical points for dynamical system describing model with matter
$\Omega_{m}\ne0$ and arbitrary coupling constant $\xi$.}
\label{tab:2}
\begin{tabular}{|c|c|c|c|c|}
\hline
& Critical point & $w_{\text{eff}}$ & eigenvalues & existence \\
\hline
1)& $x_{0}=0,y_{0}=0,\lambda_{0}=\text{const}$ & $w_{m}$ &
$l_{1}=-\frac{3}{4}(1-w_{m})(1+\sqrt{1-\frac{16}{3}\xi\frac{1-3w_{m}}{(1-w_{m})^{2}}})$
 & $z(\lambda_{0})=0$ \\
 & & & $l_{2}=\frac{3}{2}(1+w_{m})$ & \\
 & & &
 $l_{3}=-\frac{3}{4}(1-w_{m})(1-\sqrt{1-\frac{16}{3}\xi\frac{1-3w_{m}}{(1-w_{m})^{2}}})$
 &
\\
\hline
2) & $x_{0}=0,y_{0}^{2}=1,\lambda_{0}=0$ & $-1$ &
$l_{1}=-\frac{1}{2}(3+\sqrt{9+\ve2\alpha-48\xi})$ & $z(\lambda_{0})=0$\\
& & & $l_{2}=-3(1+w_{m})$ & \\
& & & $l_{3}=-\frac{1}{2}(3-\sqrt{9+\ve2\alpha-48\xi})$ &\\
\hline
3) & $x_{0}=0,y_{0}=0,\lambda_{0}=\text{const}$ &
$\frac{1}{3}$ & $l_{1}=-6\xi$ & $\ve\xi>0$\\
& & & $l_{2}=12\xi$ & $z^{2}(\lambda_{0})=\frac{1}{\ve6\xi}$\\
& & & $l_{3}=6\xi(1-3w_{m})$ &\\
\hline
\end{tabular}
\end{table}

\section{Conclusion}

In this paper we extended the analysis of the dynamics of the FRW model with 
a minimal coupling to gravity scalar field (both canonical and phantom) to the 
case of a non-minimal  coupling. We showed that in the general case the dynamics
can be represented by a four-dimensional dynamical system. However, with the 
assumption that the form of the potential parameter is $\Gamma = \Gamma(\lambda)$,
the corresponding system can be reduced to the form of a 3-dimensional dynamical 
system. Using the dynamical system methods we analyzed critical points appearing 
at a finite domain of the phase space and trajectories in their neighborhood. 
These trajectories can be obtained from the linearization of the system around 
a critical point of the type being determined by eigenvalues of the linearization 
matrix.

We found some additional points in comparison to the case of the minimal 
coupling. There are in principle at most four families of critical points which 
character (stability) depend on the value of the non-minimal coupling constant. 
In the special case of $\xi=0$ this system was investigated by Zhou
\cite{Zhou:2007xp} 
but we found that the critical points which were established by them lie rather 
at infinity ($\phi=\infty$). These critical points are out of interest because 
$\phi=\infty$ may lead to singularities. However, we use the same methodology as
Zhou and instead of postulating quintessence potential directly the relation
between $\Gamma$ and $\lambda$ is proposed.

For our dynamical analysis it would be useful to distinguish two cases: with
matter
$\Omega_{m}\ne0$ and without matter $\Omega_{m}=0$. The former is corresponding 
to a 3-dimensional dynamical system; in the latter one variable 
is eliminated due to the constraint condition $\Omega(x,y,z)=0$. For the case 
with matter we have three families of the critical points
\begin{enumerate}
\item $(0,0,0)$ --- the barotropic fluid dominated universe
which is unstable if $w_{m}>-1$ and stable if $w_{m}<-1$. 
\item $(0,\pm 1, 0)$ --- the de Sitter attractor (repellor) 
which is unstable if $w_{m}<-1$ and stable if $w_{m}>-1$.
\item $(0,0,\pm \frac{1}{\sqrt{6\ve\xi}})$ --- it is a radiation dominated 
universe of the saddle type; here are the conditions $\xi \ne 0$ and 
$\ve \xi > 0$.
\end{enumerate}
We can see that the barotropic fluid dominated universe is stable (unstable)
when the de Sitter is unstable (stable).

From the dynamical analysis of the 3-dimensional dynamical system (see
Fig.~\ref{fig:2}) it is found a generic quintessence evolution scenario (in 
the sense that it is realized by a very wide range of initial conditions). In 
this scenario the final state is represented by the de Sitter attractor with 
$w_{\text{eff}}=-1$ (see Fig.~\ref{fig:3}) or the Einstein-de Sitter universe 
with $w_{m}>-1$ and $w_{m}<-1$ respectively. Therefore all roads lead to the 
quintessence model and give the current acceleration.

Let us consider the trajectory starting from the point $(x_{0},y_{0},z_{0})$ 
then it goes to a close neighborhood of a saddle point (a radiation dominated 
universe) then approaches to stable critical point, i.e. for the $w_{m}<-1$ matter 
dominated universe or $w_{m}>-1$ de Sitter state. In the latter the trajectory 
goes close to a barotropic fluid dominated universe (unstable point) before
launch to the de Sitter state.

For the case without the matter $\Omega_{m}=0$ we have found the new generic 
quintessence scenario which appears only for non-vanishing coupling constant.
In this scenario in the phase variables (see Fig.~\ref{fig:1}) trajectories 
spend long time in the neighborhood of the saddle and then escape to the 
de Sitter attractor which leads to accelerated expansion.

\begin{acknowledgments}
MS is very grateful to prof. Antonio Zichichi and the organizers of
the International School of Subnuclear Physics 46th Course: Homage to Sidney
Coleman "Predicted and Totally Unexpected in the Energy Frontier Opened by LHC" 
held at the Ettore Majorana Foundation and Centre for Scientific Culture
(Erice-Sicily) 29 August - 7 September 2008, for hospitality where part of this paper was prepared.
We wish to thank Adam Krawiec and Pawe{\l} Tambor for useful comments and
discussion.
This work has been supported by the Marie Curie Host Fellowships for the
Transfer of Knowledge project COCOS (Contract No. MTKD-CT-2004-517186).
\end{acknowledgments}

\end{document}